# The Second Wave of the Global Crisis? A Log-Periodic Oscillation Analysis of Commodity Price Series

***Akaev, Askar A.***, Prigogine Institute of Mathematical Investigations of Complex Systems at the Moscow State University and "Complex System Analysis and Mathematical Modeling of the World Dynamics" Project, Russian Academy of Sciences; the First President of the Kyrgyz Republic

***Fomin, Alexey A.***, "Complex System Analysis and Mathematical Modeling of the World Dynamics" Project, Russian Academy of Sciences

***Korotayev, Andrey V.***, Russian State University for the Humanities, Oriental Institute and Institute for African Studies, Russian Academy of Sciences, Moscow, and "Complex System Analysis and Mathematical Modeling of the World Dynamics" Project, Russian Academy of Sciences

This article continues our analysis of the gold price dynamics that was published in December 2010 (http://arxiv.org/abs/1012.4118) and forecasted the possibility of the "burst of the gold bubble" in April – June 2011. Our recent analysis suggests the possibility of one more substantial fluctuation before the final collapse in July 2011. On the other hand, in early 2011 we detected a number of other commodity bubbles and forecasted the start of their collapse in May – June 2011. We demonstrate that this collapse has actually begun, which in conjunction with the forthcoming burst of the gold bubble suggests that the World System is entering a bifurcation zone bearing rather high risks of the second wave of the global financial-economic crisis. Indeed, on the one hand, it is obvious that such a collapse may lead to huge losses or even bankruptcies of many of the major participants of exchange games and their dependent firms and banks. Therefore, the immediate market reaction is likely to be entirely negative. Negative impact on the market could be well amplified by numerous publications in the media and business press, drawing analogies with the events of the early 1980s and earlier similar events, as well as by losses of shareholders of bankrupt companies. The current instability of major world currencies and the most powerful world economies, unfortunately, does not preclude the escalation of a short-term downswing into the second wave of economic and financial crisis. On the other hand, we have already mentioned that investments in gold, which caused the very "gold rush", also lead to the diversion of funds from stock market investments and to the reduction in the production of goods and services. If at the time of the collapse some promising areas of investment appear in the developed and / or developing countries, the investment can move to those markets, which, on the contrary, could contribute to the production of new goods and services and accelerate the way out of the crisis. It is also obvious that the decline of the oil (and other energy resources) prices may contribute very significantly to the acceleration of the world economic growth rates and the world economic recovery. The same goes for many other mineral resources whose prices have started to collapse recently, as this bares both the risk of the second wave of the world economic recession and opportunities that could stimulate the acceleration of the world industrial growth.

In a number of seminal works by Didier Sornette, Anders Johansen and their colleagues (Sornette, Sammis 1995; Sornette, Johansen 1997, 2001; Johansen, Sornette 1999, 2001; Johansen *et al.* 1996; Sornette 2004; etc.) it has been demonstrated that accelerating log-periodic oscillations superimposed over an explosive growth trend that is described with a power-law function with a singularity (or quasi-singularity) in a finite moment of time $t_C$, are observed in situations leading to crashes and catastrophes. Employing both the methodology developed by Didier Sornette and the one of the authors, already in late 2010 we indicated the possibility of the start of the second wave of the world financial-economic crisis in the second half of 2011 (http://arxiv.org/abs/1012.4118). As possible sources of respective risks one could indicate high unemployment and low consumption activity in the USA; the weakness of real estate market in this country; the problems connected with the growing budget deficit and high debt load of the USA and a number of European countries; the accelerating growth of the world prices of food commodities, energy resources and other raw materials that was causing sociopolitical tensions (see, *e.g.*, Korotayev, Zinkina 2011*a*, 2011*b*) and slow down of the economic growth. The possibilities of the further large-scale support of the economy through the monetary stimulation policies have become rather limited (or even impossible in some cases).

The high-income economies are not capable of the self-development, because they have not experienced yet adequate reforms. The debt character of the Western economy remains intact. Debts are paid out through a traditional scheme – with new debts. Bank crediting of real economy and households decreases, whereas its conditions become harder and harder. Instead of providing credits to small and medium businesses (the virtual motor of the economic upswing) banks prefer to trade in derivatives and credit default swaps that bring substantial profits. In recent months funds moved *en mass* to commodity exchanges blowing "commodity bubbles" which influenced negatively the development of the world economy and caused its instability and deceleration of its growth. Certain recovery and growth of the world economy that were observed in the recent months in the World System core are rather logical phenomena that are regularly observed after deep crises and that do not exclude the second wave of recession.

## Factors of the possible second wave of the crisis

Milton Friedman, the Noble Prize winning founder of the Monetarist school in economics, maintained that economic crises should be cured by dramatic increases in money supplies provided to the bank system (his famous "dropping money out of a helicopter"); he criticized the FRS for not doing this in the 1930s, which, according to him, resulted in the Great Depression (Friedman 1963). He seems to be correct as regards the FRS inaction after the 1929 bank crisis. On the other hand, the proverbial saying of Sir Winston Churchill, "generals always prepare to fight the last war", appears to have turned out true with respect to economists and financiers. Following the 2008 financial-economic crisis, quite in accordance with Friedman's recommendations, the USA authorities have poured into the banking system trillions of dollars; however, they have not achieved the real recovery of the American economy.

The enormous infusions of liquidity helped to save system-constituting banks; however, they have not fully resurrected the activities of credit institutions. The situation in the USA economy (and the world economy as a whole) remains unstable and indefinite, whereas negative consequences of enormous dollar emissions are rather numerous and serious.

Thus, after the 2008 financial-economic crisis the USA (but not the USA only) employed unprecedented financial stimulating measures, including $3 trillion spent in order to save the

financial system, bringing down interest rates from 5% to zero, and the rise of the budget deficit up to 11% of the GDP. Such a high level of the budget deficit was only observed in the USA during World War II (1939–1945). What is more, the FRS was also saving leading European banks by supplying them with dollars. In general, European banks received from the FRS $800 billion as loans. This way, the Federal Reserve System of the USA was actually transformed into a sort of World Reserve System.

The problem is that all those trillions of dollars are money without secure backing. On the one hand, this is somehow irresponsible on the part of the United States, because the dollar is still the main reserve currency of the world, the main currency for payments in the international trade, and that is why this influences the world economy as a whole, as well as economies of big and small countries. For example, in last 2.5 years China has issued 6 trillion yuans (approximately one trillion dollars), but the yuan is not the world reserve currency, and that is why this only creates problems for the trading partners of China. China (as many other developing countries) is ready to keep the course of exchange of its national currency as low as possible in order to preserve its competitive advantages. The European Union has also undertaken a massive infusion of money into economy.

The American authorities have been saving the financial system and economy so energetically with enormous monetary infusions that, recently, even the friendly European Union has pointed out that the pumping of the economy with uncovered money may lead to the unacceptable growth of the budget deficit and the acceleration of inflation. On the other hand, as our mathematical analysis below of the inflation bubble indicates, such a scenario is a virtually inevitable consequence of its development according to its own internal laws. On the one hand, this might look advantageous as regards the interests of the American administration (though one may doubt whether those advantages are sufficiently secure, and the respective policy is sufficiently farsighted). First of all, the USA devalues its debts through the dollar inflation and facilitates their repayment, as those debts are nominated mostly in dollars.

On the other hand, the extremely vigorous issuing of enormous amounts of money that continues for a number of years and involves numerous countries has led to an unprecedented growth of the world raw commodity prices. The abundance of the monetary mass against the background of high investment risks pushes investors to invest not in the production but rather in the acquisition of basic raw commodities. The growing demand for mineral resources was evident in the recent months, though the declining value of money was as evident. This was exacerbated by the “war of devaluations” started by the USA. The USA needed a weak dollar, the Eurozone needed a weak euro, China needed a weak yuan, and so on. The devaluation race may lead to the trade wars, which may be interpreted as one of the main risks confronting the global economy.

Against this background in recent years investors preferred to invest enormous sums in gold, insuring their savings against the risk of the devaluation of the principal currencies (and first of all, the dollar) and pushing up the gold prices. Already in December 2010 the gold price reached such a level that it did not have (taking inflation into account) since 1983 – $1430 per a troy ounce. On the 20$^{th}$ of April, 2011 it reached a new record level – $1500.

Up to May 2011 the gold price dynamics followed an explosive scenario (see Fig. 1):

**Fig. 1.** Daily gold price dynamics, May 12, 1973 – May 3, 2011, US dollars/troy ounce

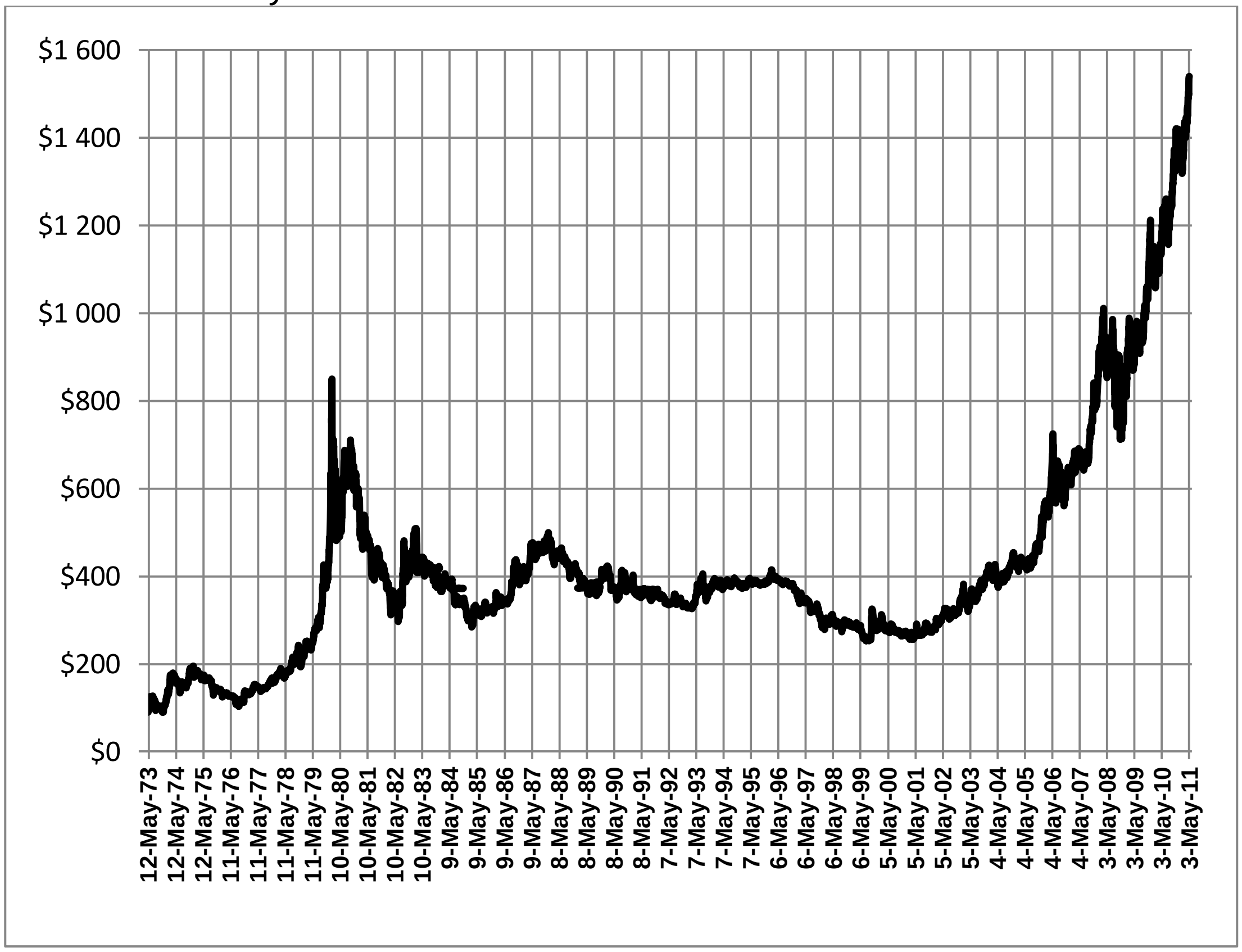

*Source*: USA Gold Reference Library database. URL: http://www.usagold.com/reference/prices/history.html.

Our earlier studies (Akaev et al. 2010; Tsirel *et al.* 2010; Akaev, Sadovnichii, Korotayev 2011) indicated that we were dealing with the blowing of a new price bubble (analogous in many respects to the one that burst at the downswing phase of the previous Kondratieff wave[1]), whereas our log-periodic oscillation analysis[2] indicated that this "gold bubble" was likely to burst in April – June 2011.

In recent months an analogous dynamics was observed with respect to the other precious metals (but not only those metals). For example, our analysis of palladium price series indicated that the palladium price bubble was supposed to collapse quite close to the gold one (Akaev, Korotayev, Fomin 2011).

In recent months one could also observe an explosive growth of the oil prices (see Fig. 2):

---

[1] On this wave see, e.g., Korotayev, Tsirel 2010.

[2] On this methodology see, *e.g.*, Akaev *et al.* 2010; Akaev, Sadovnichii, Korotayev 2011; Sornette 2004; Sornette, Johansen 1997, 1998, 2001; Sornette, Sammis 1995; Sornette, Woodard, Zhou 2009; Johansen, Sornette 1999, 2001; Johansen, Sornette, Ledoit 1999; Johansen *et al.* 1996; let us recollect that the basic equation derived by Sornette and tested on many historical examples of bubbles has the following form:

$$p(t) = A - m\,(t_c - t)^{\alpha}\,\{\,1 + C\cos[\omega \ln(t_c - t) + \varphi]\,\}, \qquad (1)$$

where $p(t)$ is the respective price at the moment $t$; $t_c$ is the "critical time"; $A$, $m$, $C$, $\alpha$, $\omega$, and $\varphi$ are constants which are to be defined on the basis of data on respective commodity prices from the start of the bubble formation till the forecast moment.

**Fig. 2.** Europe Brent daily price dynamics, June 1, 2010 – May 2, 2011, US dollars/barrel

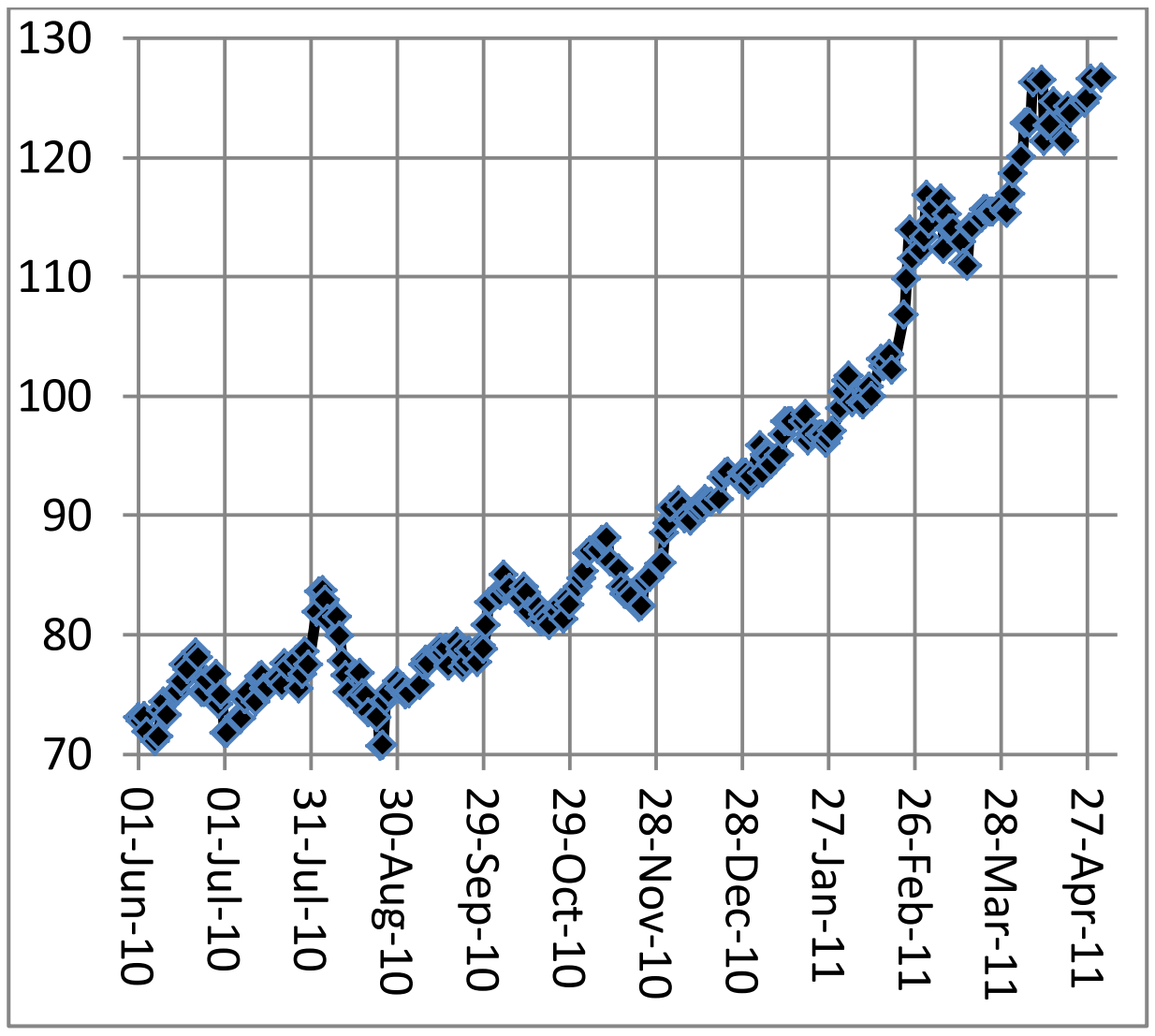

*Source*: U. S. Energy Information Administration database. URL: http://tonto.eia.gov/dnav/pet/hist/LeafHandler.ashx?n=PET&s=RBRTE&f=D.

Our earlier study performed in early 2011 (Akaev, Sadovnichii, Korotayev 2011; Akaev, Korotayev, Fomin 2011) indicated that we were also dealing with the blowing of an oil price bubble, whereas our log-periodic oscillation analysis indicated that this "oil bubble" was likely to burst in May – July 2011.

In the months preceding May 2011 one could observe an especially explosive growth with respect to the prices of silver. The analysis that we performed in mid April 2011 through the approximation of the empirical data on silver prices in the recent months with a power-law function with accelerating log-periodic oscillation superimposed over it produced the following results (see Fig. 3):

**Fig. 3.** Log-periodic oscillations in the world silver price dynamics, October 21, 2008 – April 19, 2011, , US dollars/troy ounce

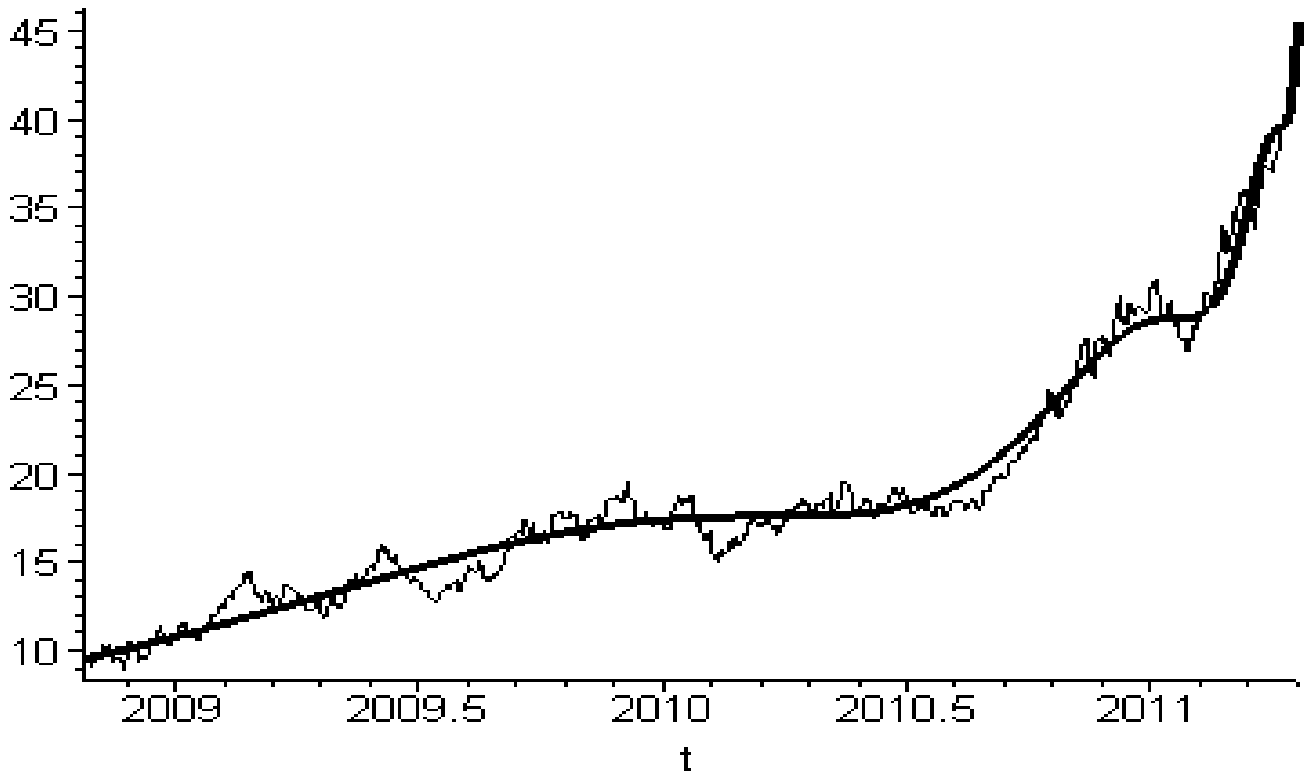

*Source of empirical data*: FOREX data base. URL: http://j64.forexpf.ru/delta/graph_popup.jsp?type=SILVER&port=.

In Fig. 3, the thin line indicates daily silver price between October 21, 2008 and April 19, 2011, whereas the smooth thick black line has been generated by the following version of equation (1) with parameters chosen by the least squares:

$$p(t) = 277.73 - 259.3\,(2011.34 - t)^{0.032}\,\{1 - 0.0069\cos[4{,}59\ln(2011.336 - t) + 4.237]\}, \quad (1a)$$

where $p(t)$ is silver price at the moment $t$.

As we see, our analysis (note that it was published more than a week before the end of April 2011 [Fomin 2011]) indicated that the burst of the "silver bubble" was most likely to happen on the 1st of May, 2011 (our respective publication of April 22, 2011 was actually titled "The Silver Bubble is Likely to Burst on the 1st of May"). Our forecast turned out to be very accurate indeed. The silver bubble burst precisely on the 1st of May, 2011 (see Fig. 4 and Table 1):

**Fig. 4.** The 1st of May burst of the "silver bubble": daily silver price dynamics, April 1 – May 21, 2011, US dollars/troy ounce

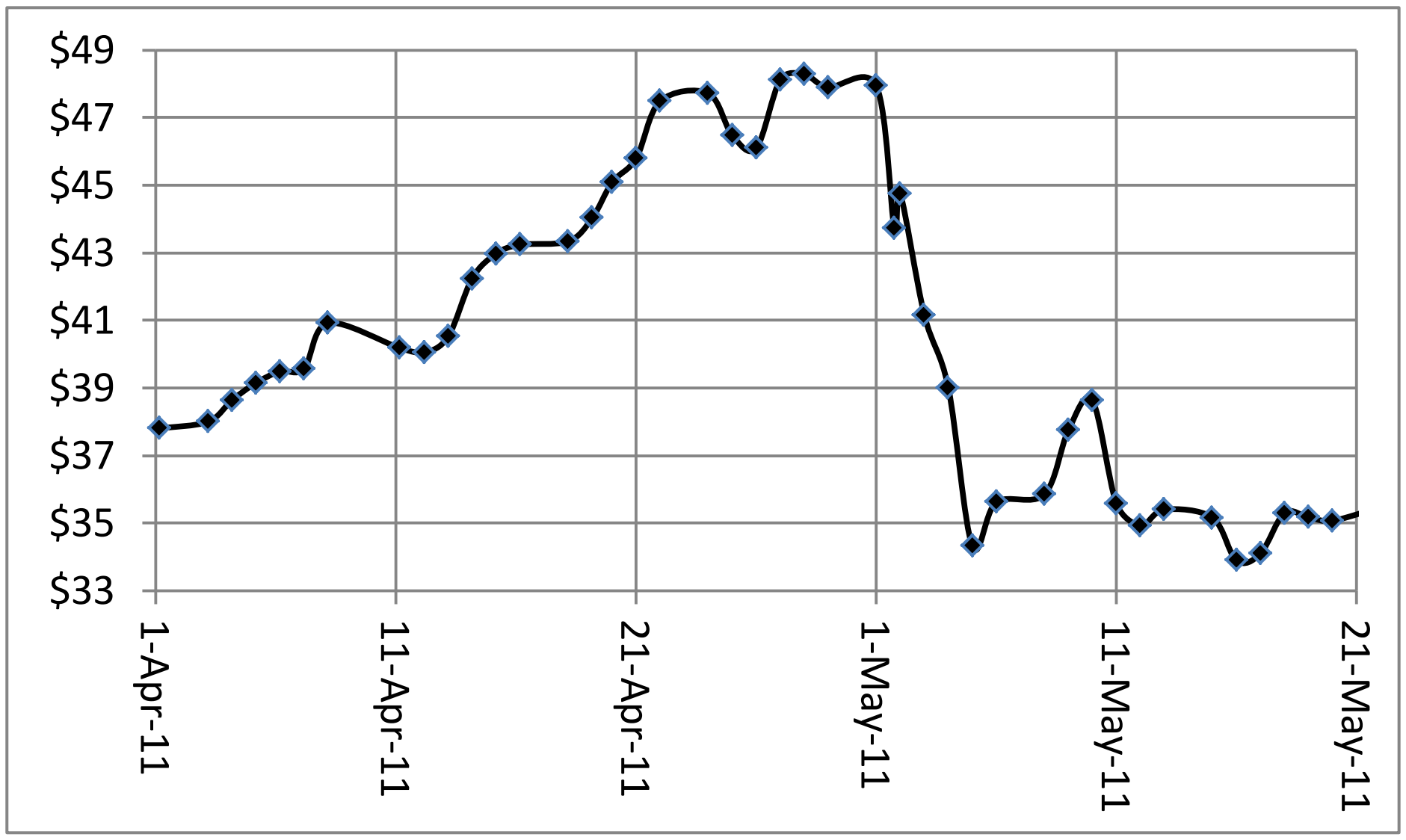

*Data source*: FOREX data base. URL: http://j64.forexpf.ru/delta/graph_popup.jsp?type=SILVER&port=.

**Table 1.** The 1st of May burst of the "silver bubble": FOREX trading system silver price dynamics, , April 28 – May 4, 2011, US dollars/troy ounce

| Date | Open | High | Low | Close |
|---|---|---|---|---|
| April 28 | 48,17 | 49,52 | 47,30 | 48,28 |
| April 29 | 48,29 | 49,16 | 47,58 | 47,90 |
| **May 1** | **47,94** | **48,15** | **42,67** | **43,72** |
| May 2 | 43,71 | 47,35 | 43,22 | 44,74 |
| May 3 | 44,77 | 45,57 | 40,60 | 41,16 |
| May 4 | 39,03 | 39,57 | 34,29 | 34,32 |

*Data source*: FOREX data base. URL: http://j64.forexpf.ru/delta/graph_popup.jsp? type=SILVER&port=.

As we remember, our earlier analysis suggested that the burst of the gold and oil bubbles should have happened soon after the burst of the silver bubble. Indeed, as regards the oil prices, since early May 2011 we seem to have been observing the formation of an oil price antibubble (see Figs. 5 and 6):

**Fig. 5.** Europe Brent daily price dynamics,
August 10, 2010 – June 30, 2011, US dollars/barrel

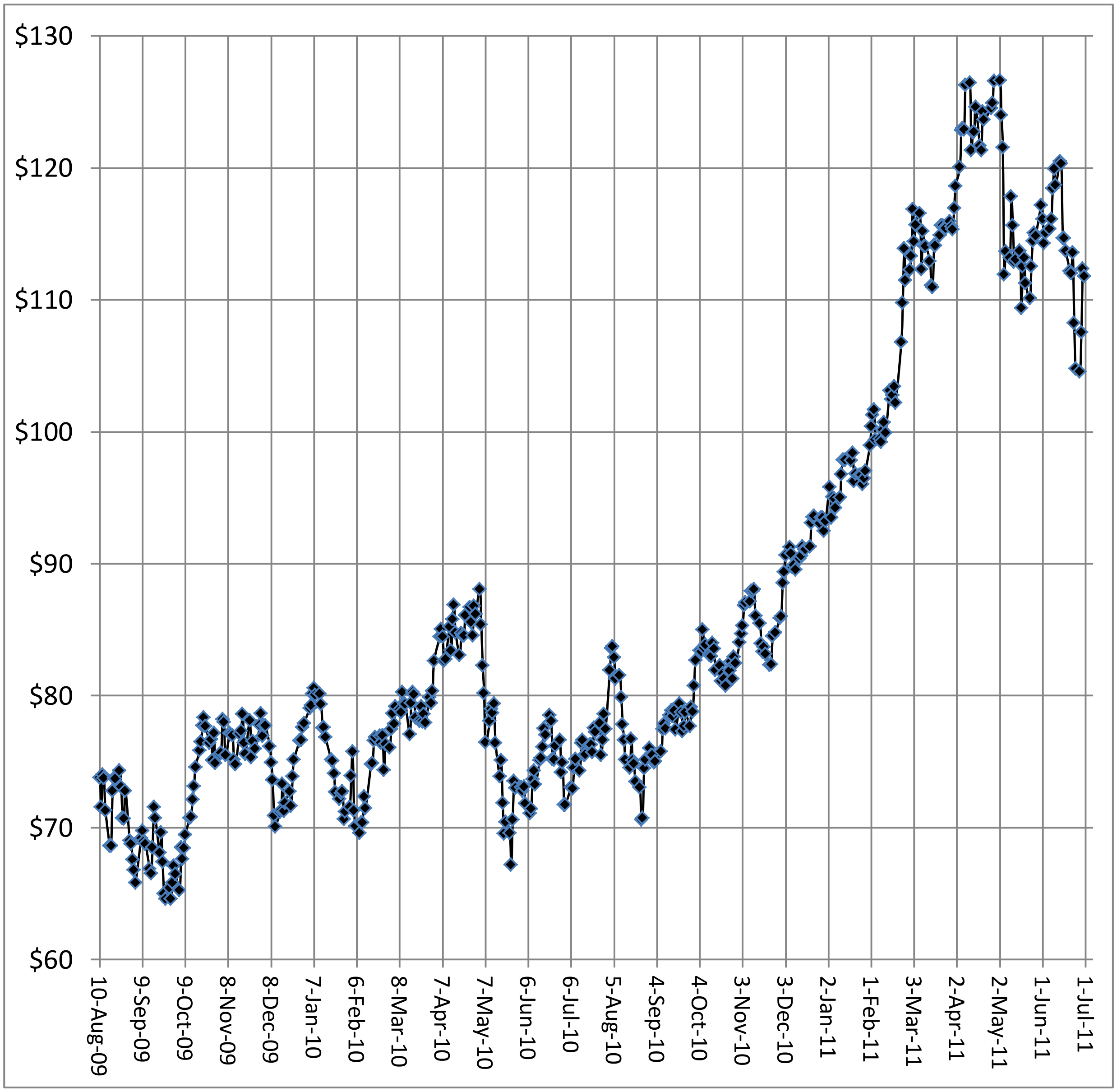

*Data sources*: U. S. Energy Information Administration database. URL: http://tonto.eia.gov/dnav/pet/hist/LeafHandler.ashx?n=PET&s=RBRTE&f=D (August 10, 2009 – June 28, 2011); FOREX data base. URL: http://j64.forexpf.ru/delta/graph_popup.jsp?type=BRENT&port= (June 29 and 30, 2011).

**Fig. 6.** Formation of an "oil antibubble" since early May, 2011?
Europe Brent daily price dynamics,
January 12, 2011 – June 30, 2011, US dollars/barrel

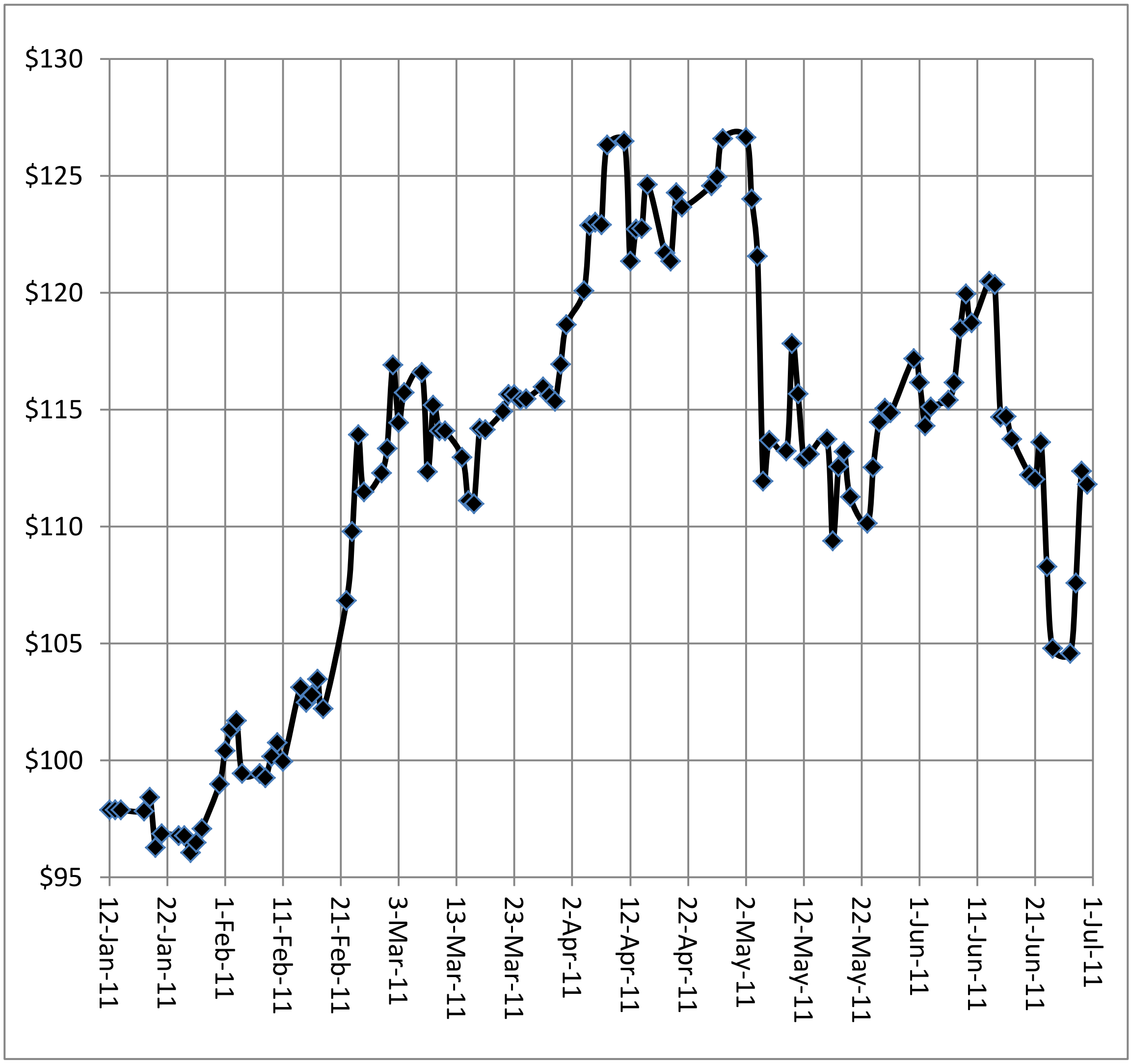

*Data sources*: U. S. Energy Information Administration database. URL: http://tonto.eia.gov/ dnav/pet/hist/ LeafHandler.ashx?n=PET&s=RBRTE&f=D (January 12, 2010 – June 28, 2011); FOREX data base. URL: http://j64.forexpf.ru/delta/graph_popup.jsp? type=BRENT&port= (June 29 and 30, 2011).

As regards the gold price dynamics, it does not appear possible to say for sure that the gold "antibubble" has already started its formation (though this cannot be excluded entirely either) (see Fig. 7):

**Fig. 7.** Daily gold price dynamics, February 7, 2010 – June 28, 2011, US dollars/troy ounce

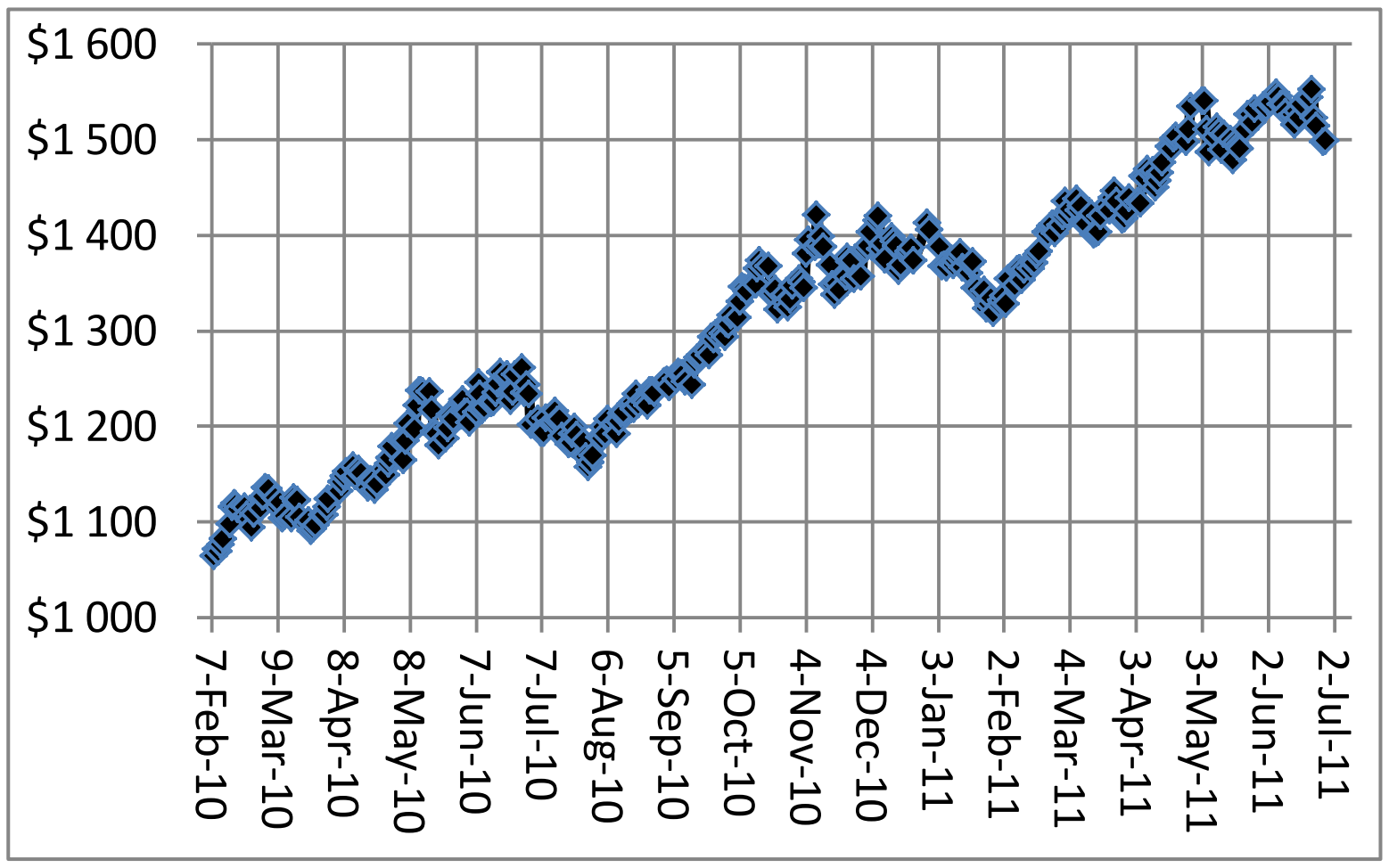

*Source*: USA Gold Reference Library database. URL: http://www.usagold.com/reference/prices/history.html.

Indeed, as our analysis of the respective time series including more recent data suggests, there are no grounds to exclude the possibility of one more substantial oscillation before the final collapse of the gold bubble (see Fig. 8):

**Fig. 8.** Log-periodic oscillations in the world gold price dynamics, November 8, 2003 – May 26, 2011, inflation adjusted March 2011 US dollars/troy ounce

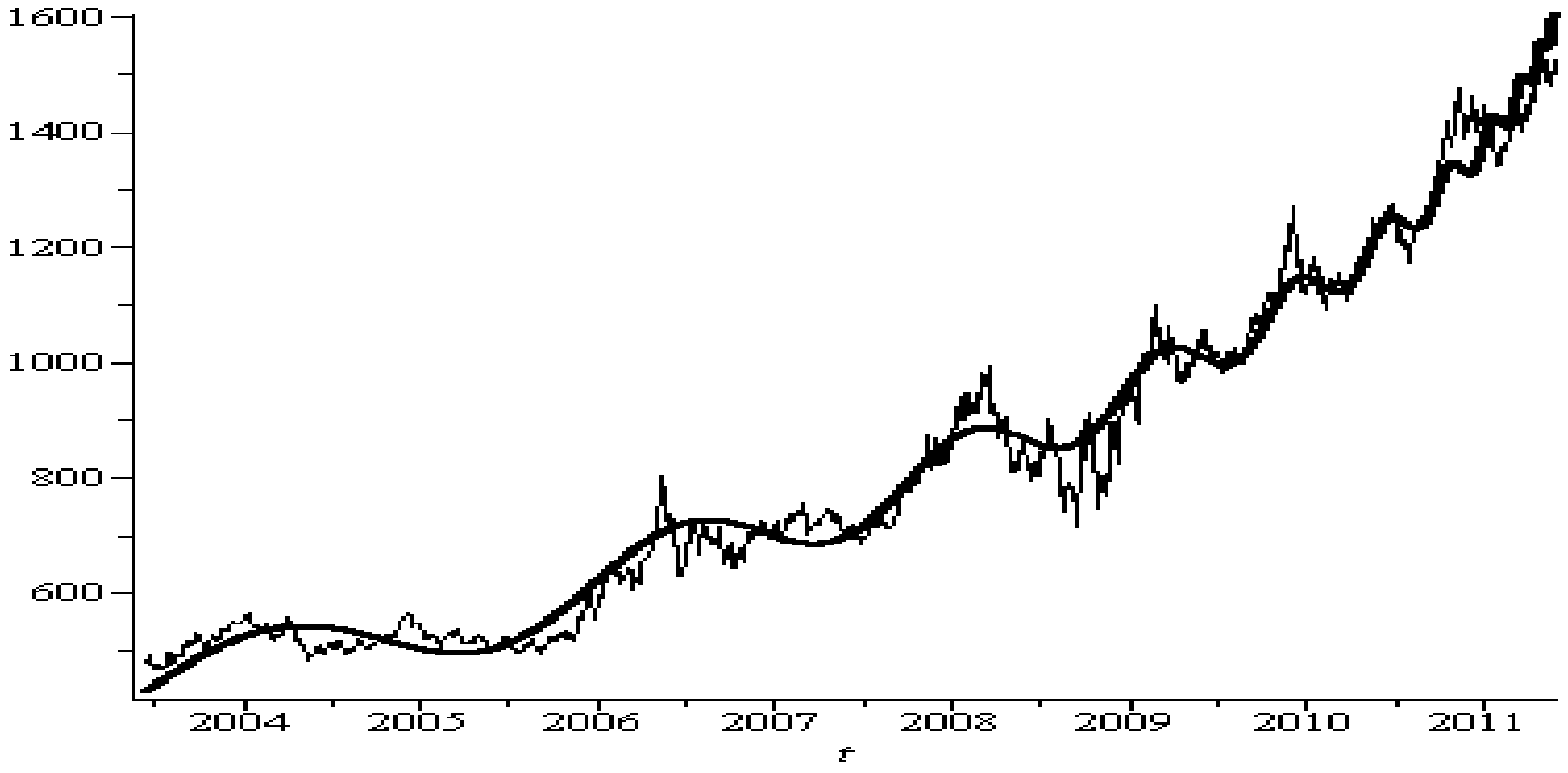

*Source of empirical data*: USA Gold Reference Library database. URL: http://www.usagold. com/ reference/prices/ history.html.

In Fig. 8, the thin line indicates daily gold price between November 3, 2003 and May 26, 2011, whereas the smooth thick black line has been generated by the following version of equation (1) with parameters chosen by the least squares:

$$p(\mathrm{t}) = 1978.2 - 734.8\,(2011.573 - t)0.36\,\{1 + 0.024\cos[16.5\ln(2011.573 - t) - 36.3]\}, \qquad (1b)$$

where $p(t)$ is gold price at the moment $t$.

Note that the quasisingularity moment ($t_C$) here equals 2011,573, which corresponds to July 27 and suggests that the gold bubble should start collapsing before this date anyway.

Hence some of the commodity price bubbles that were blown up in the recent months have already started collapsing, whereas the other are likely to collapse in the forthcoming weeks. As we have already indicated in our earlier publications, this implies a strong risk of the forthcoming second wave of the world financial-economic crisis. Indeed, on the one hand, it is obvious that such a collapse may lead to huge losses or even bankruptcies of many of the major participants of exchange games and their dependent firms and banks. Therefore, the immediate market reaction is likely to be entirely negative. Negative impact on the market may well be amplified by numerous publications in the media and business press, drawing analogies with the events of the early 1980s and earlier similar events, as well as by losses of shareholders of bankrupt companies. The current instability of major world currencies and the most powerful world economies, unfortunately, does not preclude the escalation of a short-term downswing into the second wave of economic and financial crisis.

On the other hand, we have already mentioned in our earlier publications that investments in gold, which caused the very "gold rush", also lead to the diversion of funds from stock market investments and to the reduction in the production of goods and services. If at the time of the collapse some promising areas of investment appear in the developed and / or developing countries, the investment can move to those markets, which, on the contrary, may contribute to the production of new goods and services and accelerate the way out of the crisis. It is also obvious that the decline of the prices of oil (and other energy resources) could contribute very significantly to the acceleration of the world economic growth rates and the world economic recovery. The same goes for many other mineral resources whose prices have started to collapse recently, as this bares both the risk of the second wave of the world economic recession and opportunities that could stimulate the acceleration of the world industrial growth (see, *e.g.*, Fig. 9):

**Fig. 9.**FOREX trading system aluminium price dynamics, May 2010 – June 2011, US dollars/metric ton

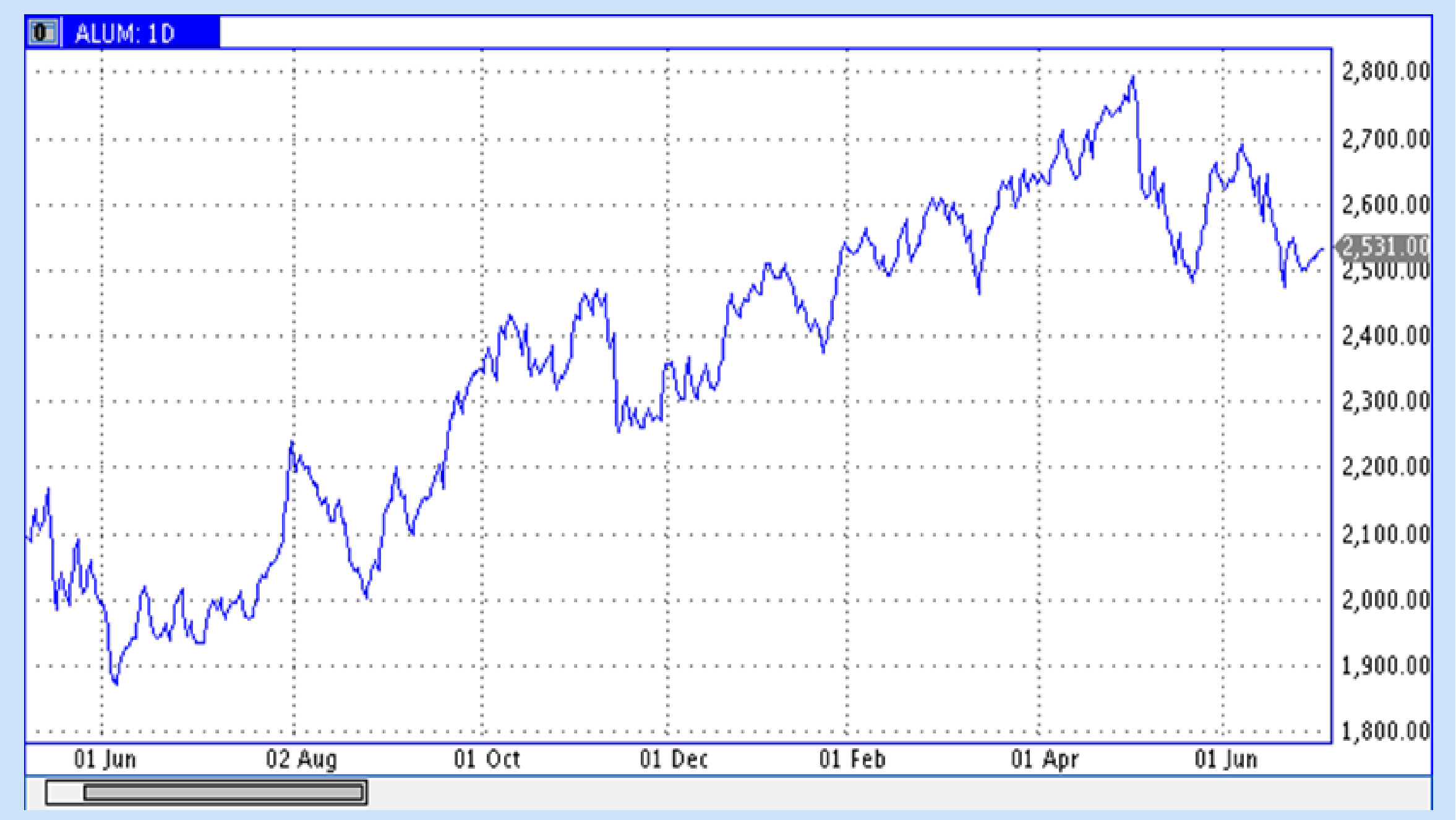

*Data source*: FOREX data base. URL: http://www.forexpf.ru/chart/aluminium/.

Thus, in many respects the World System has just entered an actual bifurcation zone that implies both the risks of the second wave of the world recession and the possibilities to achieve a steady world economic recovery and serious acceleration of the world economic growth rates. As in any bifurcation point very much will depend on the actions of the main relevant actors.

* * *

Unfortunately, there are a number of factors that make the pessimistic scenario rather likely and that will be considered in detail in our next paper.